\documentclass[a4paper,11pt]{article}
\pdfoutput=1 
\usepackage{jheppub} 
\usepackage[]{epsfig}
\usepackage{amsfonts}
\usepackage{amsmath,bm}
\usepackage{subfigure}
\usepackage{graphicx}
\usepackage{graphics}
\usepackage{bbold}
\usepackage{slashed}

\def\ba{\begin{eqnarray}}
\def\ea{\end{eqnarray}}
\def\be{\begin{equation}}
\def\ee{\end{equation}}

\renewcommand{\Im}[1]{\text{Im}{#1}}
\title{Probing holographic flat bands at finite density}

\author[a,b]{Nicol\'as Grandi}\author[c,d]{Vladimir Juri\v ci\' c}
\author[a]{Ignacio Salazar Landea}
\author[e]{Rodrigo Soto-Garrido}

\affiliation[a]{Instituto de F\'\i sica de La Plata - CONICET, C.C. 67, 1900 La Plata, Argentina}
\affiliation[b]{Departamento de F\'\i sica - UNLP, Calle 49 y 115 s/n, 1900 La Plata, Argentina}

\affiliation[c]{Departamento de F\'isica, Universidad T\'ecnica Federico Santa Mar\'ia, Casilla 110, Valpara\'iso, Chile.}

\affiliation[d]{Nordita, KTH Royal Institute of Technology and Stockholm University, Roslagstullsbacken 23, 10691 Stockholm, Sweden}

\affiliation[e]{Facultad de F\'isica, Pontificia Universidad Cat\'olica de Chile, Vicu\~{n}a Mackenna 4860, Santiago, Chile}

\emailAdd{grandi@fisica.unlp.edu.ar}
\emailAdd{juricic@gmail.com}
\emailAdd{peznacho@gmail.com}
\emailAdd{rodrigo.sotog@gmail.com}

\abstract{Flat band electronic systems exhibit a rich landscape of correlation-driven phases, both at the charge neutrality and finite electronic density, featuring exotic electromagnetic and thermodynamic responses. Motivated by these developments, in this paper,  we explicitly include the effects of the chemical potential in a holographic model featuring approximately flat bands. In particular, we explore the  phase diagram of this holographic flat band system  as a function of the chemical potential. We find that at low temperatures and densities,  the system features a  nematic phase, transitioning into the Lifshitz phase as the chemical potential or temperature increases. To further characterize the ensuing phases, we investigate the optical conductivity and find that this observable shows strong anisotropies in the nematic phase. 
}

\begin{document}
\maketitle

\section{Introduction}

Condensed-matter  systems featuring  flat electronic bands  have drawn significant attention in both theoretical and experimental condensed matter communities  since their experimental realization in the magic angle twisted bilayer graphene~\cite{cao2018correlated,cao2018unconventional}. 
Due to their flatness, \emph{i.e.} quenching of the electrons' kinetic energy, strong electron-electron correlations dominate in these systems yielding a broad landscape of quantum phases~\cite{andrei2020graphene}. Furthermore,  by tuning external parameters, such as doping~\cite{cao2018unconventional,lu2019,yankowitz2019,sharpe2019,Zondiner2020} and strain~\cite{Zhang2022}, the prospect of realizing correlation-driven phases in flat band systems is further enriched. 

\vspace{1mm}
 
To address the effects of the interactions in flat bands, non-perturbative means are therefore desirable, 
motivating the use of the AdS/CFT holographic duality \cite{Zaanen-2015,hartnoll2018,Zaanen:2021}.
Several holographic models have been so far proposed, where flat bands are manifest and yield distinct phase diagrams~\cite{Laia:2011zn,Grandi2021,Grandi2022,seo2022}. Previous works on other holographic realizations of topological phases of matter include Refs.~ \cite{Landsteiner:2015lsa,Liu:2018bye,Dantas:2019rgp,Juricic:2020sgg,BitaghsirFadafan:2020lkh,Ji:2021aan}.
In particular, in the model introduced in Refs.~\cite{Grandi2021,Grandi2022},  the emerging Lifshitz geometry gives rise to the  flattening of the bands by tuning the dynamical exponent $z$. In turn, the flattened bands can be probed by explicitly including fermions in the model \cite{Grandi2022}, yielding a dispersion relation 
for the fermionic degrees of freedom 
 \begin{equation}
     \omega \approx \pm D k^2-i\Gamma k^4,
     \label{quadbandtouch}
 \end{equation}
with the parameters $D,\Gamma>0$.  We emphasize that such a  dispersion relation features a vanishing Fermi velocity, and  can thus be considered as a flattened version of the usual linear Dirac dispersion~\cite{bistritzer2011moire}. Besides the (rotationally invariant) Lifshitz phase, this holographic  model  also features a nematic (rotational-symmetry breaking) phase, 
with a temperature-driven transition between these phases~\cite{Grandi2021}. 
 
\vspace{1mm}

The chemical potential, however,   has not been incorporated in the original model~\cite{Grandi2021}, which should be of a crucial  importance for  understanding  the fate of a flat band away from charge neutrality, as expected based on the fact that the density of states is then enhanced, and should therefore make the system even more prone to the effects of the electronic interactions. Such   a situation is also quite natural from an experimental point of view, where a plethora  of competing orders was observed, for instance,  in twisted bilayer graphene~\cite{cao2018correlated,cao2018unconventional,lu2019,yankowitz2019,sharpe2019,Zondiner2020,Zhang2022,andrei2020graphene}. 

To answer this rather important question, which is  both theoretically and the experimentally consequential, here we explicitly include the chemical potential in the model, thereby addressing the effects pertaining to   finite  density of fermions. In that respect, it is of a particular importance to consider responses of the ensuing phases in such a setup. Motivated by this, we analyze the response of this system to the electromagnetic field by computing  the optical conductivity. 

\vspace{.2cm}
\paragraph{Key results:} We here show that the chemical potential drives the holographic flat bands from a low-density nematic phase to a high-density Lifshitz 
phase, with the obtained phase diagram at finite chemical potential and temperature shown in Fig.~\ref{fig:phase-diagram}.  Furthermore, we find that the conductivity at finite temperature  in the nematic phase is anisotropic, with the anisotropy that gradually disappears as the system approaches the Lifshitz phase, as shown in Fig.~\ref{fig:DCconductivity}. This is further corroborated by the form of the 
conductivity in the zero-frequency limit,
which shows a very strong anisotropy in the nematic phase and smoothens out as 
the system approaches the phase transition, see Fig.~\ref{fig:ACconductivities1}. 
 
\vspace{3mm}
\section{Model}

The holographic model for flat bands introduced in Ref.~\cite{Grandi2021} is defined by the bulk gravitational action
\begin{equation}
S= 
\int d^4x\sqrt{-g}\left(R-2\Lambda\right)
-\frac{1}{4}\int
\left[ F\wedge {^{\!\star}  F} +
\mathrm{Tr}\left(
G\wedge{^{\!\star} G}\right)
\right],
\label{eq:HoloAct}
\end{equation}
with the first term corresponding to the standard Einstein-Hilbert dynamics, while in the second  ${F}=d A$ represents a $U(1)$ gauge field strength accounting for the conservation of particle number on the boundary theory. Furthermore,   $G=d{B} - i(q/2) {B}\wedge B$ is the strength of an $SU(2)$ gauge field representing a UV flavor symmetry. This can be interpreted as the emergent  low energy description of a bilayer structure in the boundary theory.

To switch on a finite temperature, we search for  asymptotically AdS black hole solutions, with the generic \emph{ansatz} for the metric
\begin{equation}
\mathrm d s^2 = \frac{1}{r^2}\Big(\!-\!N(r) f(r)\mathrm dt^2 + \frac{dr^2}{f(r)}+ dx^2 + dy^2+ 
2 h(r)\, dx\, dy \Big)\,,
\label{eq:metrica}
\end{equation} 
in terms of purely $r$-dependent functions $f,N$ and $h$, which close to the boundary satisfy $f,N\to 1$ and $h\to  0$. Black hole solutions feature a  horizon at finite $r=r_h$, where $N$ and $h$ are bounded, and $f$ vanishes linearly with $f'= {4 \pi T}/{\sqrt{N}}$, with the parameter $T$ representing the boundary temperature. 

For the non-Abelian gauge field, we write
\begin{equation}
B =  \frac12\left(Q_{1}(r) \sigma_1\,+Q_{2}(r) \sigma_2\,\right) dx +\frac12  \left(Q_{1}(r)\sigma_2+Q_{2}(r)\sigma_1 \right)dy,
 \label{eq:nonAbelian}
\end{equation}
where the $\sigma_i$'s represent the standard Pauli matrices and $Q_1(r)$ and $Q_2(r)$ are both $r$-dependent functions which are finite at the horizon. On the boundary, we turn on a deformation   $Q_1\to 2m_*$, which breaks the $SU(2)$ flavor symmetry  explicitly on the dual theory side leaving it invariant  only under a combination of flavor $U(1)$ and rotations. The resulting phase diagram features a Lifshitz phase at high temperatures (with the dynamical critical exponent $z$ completely fixed by the coupling $q$), characterized by trivial solutions for $Q_2(r)=h(r)=0$. As temperature is lowered, these fields  develop  spontaneously a profile which signals a nematic phase as the temperature is lowered \cite{Grandi2021}. The quasiparticle dispersion relation is quadratic in the Lifshitz phase, and strongly anisotropic (splitting into two separated Dirac cones) in the nematic phase \cite{Grandi2022}. Hence the phase transition can be interpreted as two Berry monopoles with charge one merging into a Berry monopole with charge two.

\vspace{.2cm}
\section{Phase diagram at finite  density}
To move away from charge neutrality, we turn on the Abelian gauge field as
\begin{equation}
{A} =A_t(r) dt.\\
\end{equation}
This field is coupled to the particle current in the dual theory, the boundary value of its time-like component representing a finite chemical potential $A_t\to \mu$. In addition, to avoid conical singularities, we set $A_t\to 0$ at the horizon. {One may then be tempted to turn on an electric field in the $U(1)_3$ inside the $SU(2)$ as it has been done for some p-wave superconducting models \cite{Gubser:2008wv}. In the interpretation of $SU(2)$ as a symmetry between the (graphene) layers, this would be equivalent to a charge imbalance between the two layers. Hence, we find it more natural to turn on an overall $U(1)$ field. Notice that a $U(1)_3$ chemical potential is expected to enhance the nematic order since  the nematic order can be interpreted as an excitonic condensate.}

\begin{figure}[t!]
\centering
\includegraphics[width=0.67\textwidth]{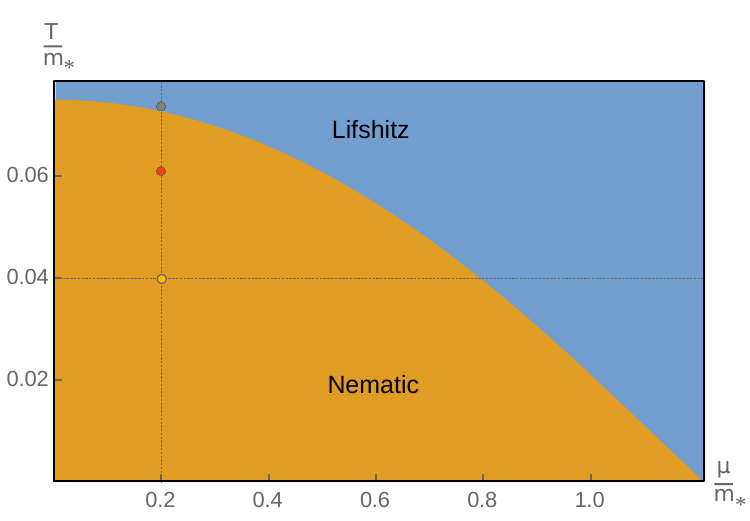} 
\vspace{.1cm}
\caption{\label{pt} Phase diagram of the  holographic flat-band system defined by Eqs.~\eqref{eq:HoloAct}-\eqref{eq:nonAbelian} in  the $\mu-T$ plane.  Observe that the nematic phase, stable at low temperatures, is rendered unstable as the chemical potential is increased to the isotropic Lifshitz one. The vertical and horizontal dotted lines correspond to the trajectories of the DC conductivity in Fig.  \ref{fig:DCconductivity}. The three colored dots mark the points for which the optical conductivity is shown  in    Fig. \ref{fig:ACconductivities1}. 
\vspace{.2cm}}
\label{fig:phase-diagram}
\end{figure}

Figure \ref{pt} shows the resulting phase diagram.
Since the quadratic band touching causes an increased density of states at zero chemical potential, we anticipate a significant chenge in the phase diagram's structure when the bands are filled, { i.e. as the chemical potential increases.} In fact,  the effect is rather mild for low enough chemical potentials as the slope of the transition curve is then equal to zero at $\mu=0$. On the other hand, as  the chemical potential is increased the nematic order gets suppressed, and  a quantum phase transition from the nematic to the flat-band or symmetric phase takes place  at a finite critical chemical potential  $\mu/m_*\approx 1.2$. This could be expected, as the energy scale associated with the chemical potential should drive the quantum fluctuations. Therefore at a sufficiently strong chemical potential, the associated quantum fluctuations wash out the anisotropy characterizing  the nematic, and the system enters the symmetric phase. Furthermore, an explicit  analysis  
of the free energy and the entropy of the system  as a function of  the temperature and chemical potential reveals that the phase transition is of the second order (continuous).

The form of the phase diagram may be understood {by making an analogy with the corresponding weakly coupled system,} as follows. Firstly, our nematic order is neutral under the global $U(1)$ symmetry. Therefore, the corresponding condensate can be associated with electron-hole pairs,  forming an excitonic condensate 
\cite{fuhrer2016,kogar2017,wang2019}. Excitonic condensates are usually short lived and a stabilizing mechanism should arise so that the electron and hole do not annihilate (recombine). In our case, the electron and hole can be considered to belong to different layers, implying  that the annihilation rate could be suppressed. This picture may also explain why a finite chemical potential disfavors the nematic phase: As the electron band is populated, hole states become less accessible, and therefore it becomes more difficult to establish a condensate. This picture is consistent  with previously reported results at weak coupling for bilayer graphene \cite{vafek2014superconductivity,murray2014excitonic} and these in the context  of brane constructions of holographic bilayer excitonic  condensates~\cite{Grignani:2014tfa,Evans:2014mva}. {
This intuition based on the  weak coupling scenario may survive even in situations where the system's excitations could be complicated composite fields \cite{hartnoll2018}.  It is, however, worth pointing out that, even though it may be helpful to provide  such an interpretation, this is not necessary.}

\begin{figure}[t!]
\centering
\includegraphics[width=0.495\textwidth]{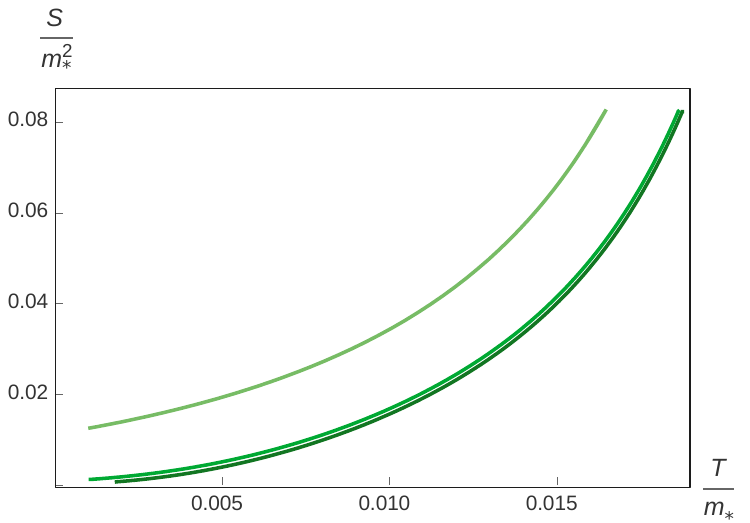} \hfill \includegraphics[width=0.495\textwidth]{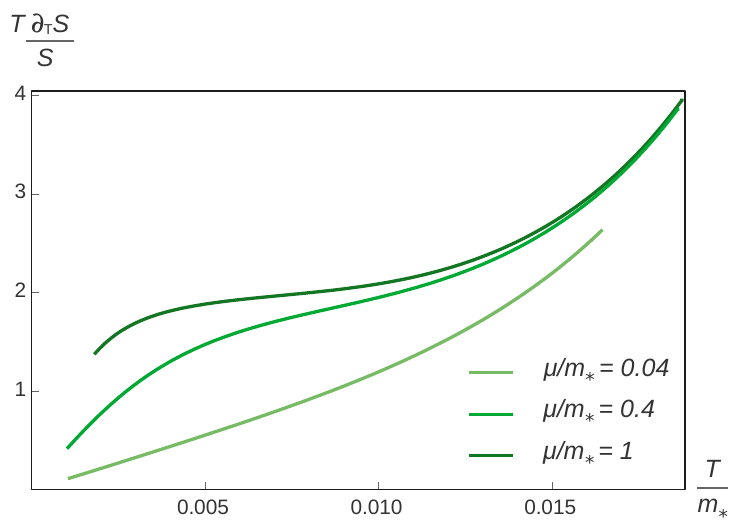} 
\put (-203,113) {(b)}
\put (-413,113) {(a)}
\vspace{.1cm}
\caption{(a) Re-scaled entropy density and (b) its corresponding logarithmic derivative with respect to the temperature, for different values of $\mu/m_*$ 
}
\label{fig:scalings}
\end{figure}

{

Let us now comment on the expected IR geometry at very low temperatures for the broken symmetry phase. A first thing to notice is that as we lower the temperature,  the values of purely radial functions $Q_i(r)$ ($i=1,2$, see eq.~\eqref{eq:nonAbelian}) at the horizon  approach one another, $Q_1(r_h)\approx Q_2(r_h)$. This implies that near the horizon the Yang-Mills curvature vanishes. The anisotropic lapse function $h(r)$ hence tends to a constant which can be scaled out at fixed $r$ to recover rotational invariance. So far this goes in parallel with the neutral case \cite{Grandi:2021bsp}, where not only rotational invariance is recovered but also full Lorentz invariance. This implies that in the deep IR one would recover an AdS$_4$ solution with the same AdS radius as in the UV. We conjecture that this represents  a hallmark feature  for holographic states of matter exhibiting Dirac cones separated in momentum space (see also Ref.~\cite{Landsteiner:2015pdh}). For charged solutions, a horizon is necessary to hide the charge  
as there are no any charged-matter field sectors available to source the electric field. We therefore  expect that the IR geometry will correspond to AdS$_2\times $R$_2$ space time, as for a Reissner-Nordström black hole.

To further understand this, we plot in Fig.~\ref{fig:scalings}(a) the entropy density defined by the re-scaled area of the horizon
\begin{equation}
    S=1-h(r_h)^2
\end{equation}
as a function of the temperature for different values of the chemical potential. From the plot we observe that the entropy tends to a chemical-potential-dependent value that increases with the chemical potential. This agrees with the idea of a ground-state characterized by a AdS$_2\times $R$_2$ geometry.

In Fig.~\ref{fig:scalings}(b) we also show the scaling of the entropy with respect to the temperature by considering the combination $\frac{T S'}S$, with $S'$ as the derivative of the entropy with respect to the temperature. Close to the phase transition, $S'$ diverges as expected for second order phase transitions, while interesting features occur as we lower the temperature. We see that for low enough values of the chemical potential we recover an intermediate regime compatible with full Lorentz invariance. This intermediate regime is likely to be characterized by an AdS$_4$ geometry as in the $\mu=0$ case. As we keep lowering the temperature eventually the scaling is lost and we flow towards a scaling compatible with $AdS_2\times R_2$. 

The degeneracy of the entropy at low temperatures [Fig.~\ref{fig:scalings}(a)] is a benchmark of $AdS_2\times R_2$ geometry, and it is supposed to be lifted by phase transitions when considering additional degrees of freedom. In fact, such geometry is known to induce superconducting phase transitions when charged operators are allowed in the dual theory \cite{Gubser:2008px}. Hence, our minimal model can provide a geometric explanation for the appearance of superconducting phases in the proximity of neutral orders~\cite{Sachdev-Vojta-PRL1999,Sachdev-RMP2003,vafek2014superconductivity,murray2014excitonic,Roy-Juricic-PRB2019}.

}



~

\section{Probing the holographic phases by conductivity} Let us probe our solutions by computing the optical conductivity. As rotation symmetry is broken, the conductivity becomes direction-dependent, with the principal axes expected to be  given by the corresponding  solution~\cite{Ji:2022ovs}, which in our case represents a nematic phase. The \emph{Ansatz} for the fluctuations therefore reads
\small
\begin{align}
    \delta B=& 
    \frac{e^{-i \omega t}}{\sqrt{2}}
    \left[
    \left(
    (a_+^{(3)}+a_-^{(3)})\,dx
    + 
    (a_+^{(3)}-a_-^{(3)})\,dy
    \right)\sigma_3  \nonumber
    + 
    \left(
    (a_t^{(+)}+a_t^{(-)}) \sigma_1
    +
    (a_t^{(+)}-a_t^{(-)}) \sigma_2
    \right)dt 
    \right]
    \\
    \delta A=&\frac{e^{-i \omega t}}{\sqrt2}\!
    \left(
    (a_+^{(0)}\!+\!a_-^{(0)})\, dx
    +
    (a_+^{(0)}\!-\!a_-^{(0)})\, dy 
    \right)
    \quad
    \delta ds^2 = 
        \frac{e^{-i \omega t}}{\sqrt{2}}\!
    \left(
    (g_{t+}\!+\!g_{t-})\,dx\,dt+ (g_{t+}\!-\!g_{t-})\,dy\,dt
    \right),
\end{align}
\normalsize 
where the $\pm$ signs correspond to the $x_\pm=x\pm y$ directions respectively.
This \emph{Ansatz} decouples the fluctuations in two sectors, one of which is composed by the fields $(a_+^{(0)},$ $a_-^{(3)},$ $a_t^{(+)},$ $g_{t+})$ and the other is given by $(a_-^{(0)},$ $ a_+^{(3)},$ $a_t^{(-)},$ $g_{t-})$, which are  functions of the  $r$ coordinate.  

Now by integrating the corresponding equations of motion with in-going boundary conditions at the horizon, we can read the optical conductivity from the leading 
and sub-leading  behavior of the fields near the AdS boundary  and obtain the corresponding expectation values by evaluating the on-shell action \cite{Hartnoll:2009sz,Kim:2014bza}. 
{Defining the electric $\langle J_\pm\rangle$ and heat $\langle Q_\pm\rangle$ currents, we can relate them with the thermal gradient $\nabla_\pm T$ and the electric field $E_\pm$ at the boundary as}
\begin{align}
       \begin{pmatrix}
    \langle J_\pm\rangle\\
    \langle Q_\pm\rangle
    \end{pmatrix}=
    \begin{pmatrix}
    \sigma & \alpha T\\
    \bar{\alpha} T & \kappa
    \end{pmatrix}
    \begin{pmatrix}
     E_\pm\\
    -(\nabla_\pm T )/T
     \end{pmatrix}
     =\begin{pmatrix}
     \sigma & \alpha T\\
     \bar{\alpha} T & \kappa
     \end{pmatrix}
    \begin{pmatrix}
    i\omega( a^{(0)}_{\pm(0)}+\mu  g_{t\pm(0)})\\
   i\omega\mu g_{t\pm(0)}
   \end{pmatrix}.
\end{align} 
{Notice that in the second equality we relate the sources with the boundary value of the bulk field perturbations in the standard way.} 
In this analysis we are omitting the propagators involving $a_\pm^{(3)}$ and $a_t^{(\mp)}$ as they are not relevant for the present analysis.

\begin{figure*}[t!]
\begin{center}
\includegraphics[width=0.495\textwidth]{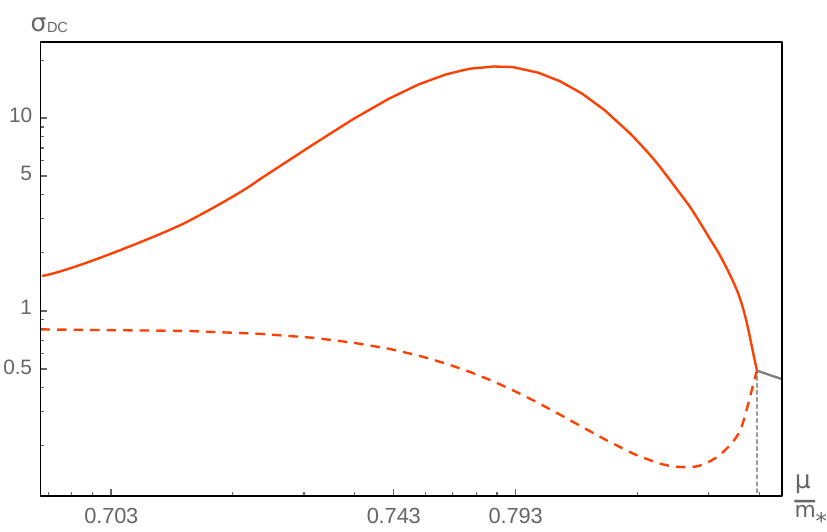}
\includegraphics[width=0.495\textwidth]{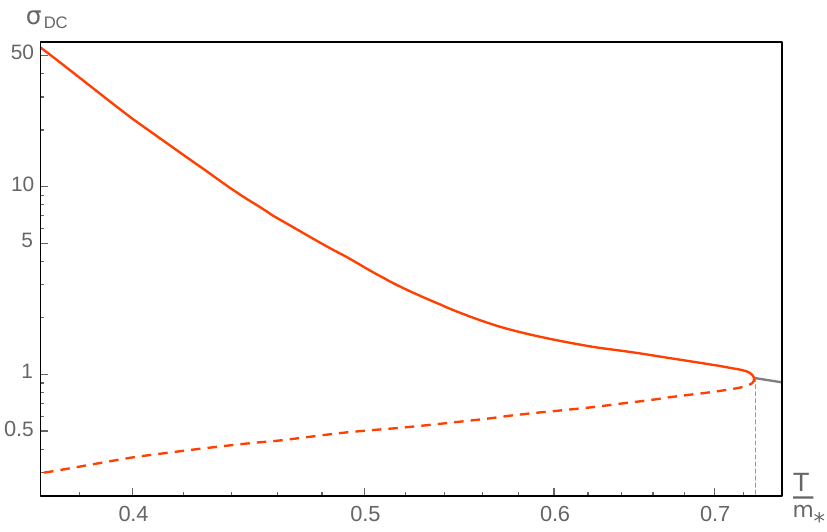}
\put (-28,115) {(b)}
\put (-417,115) {(a)}
\caption{(a) Real part of the electrical conductivity as $\omega\to0$ in the direction $x_+=x+ y$ ($x_-=x-y$) shown in solid (dashed), as a function of the chemical potential for fixed temperature $T/m_*\approx 0.04$. (b) Real part of the electrical conductivity as $\omega\to0$ in the direction $x_+=x+ y$ ($x_-=x-y$) shown in the solid (dashed) line, as a function of the temperature at fixed chemical potential $\mu/m_*\approx 0.02$. Notice that precisely at the quantum-critical point for the nematic - flat band phase transition the conductivities in the two directions become equal, and therefore the flat-band phase shows an isotropic behavior. }
\label{fig:DCconductivity}
\end{center}
\end{figure*}

\begin{figure*}[t!]
 \begin{center}
 \includegraphics[width=1\textwidth]{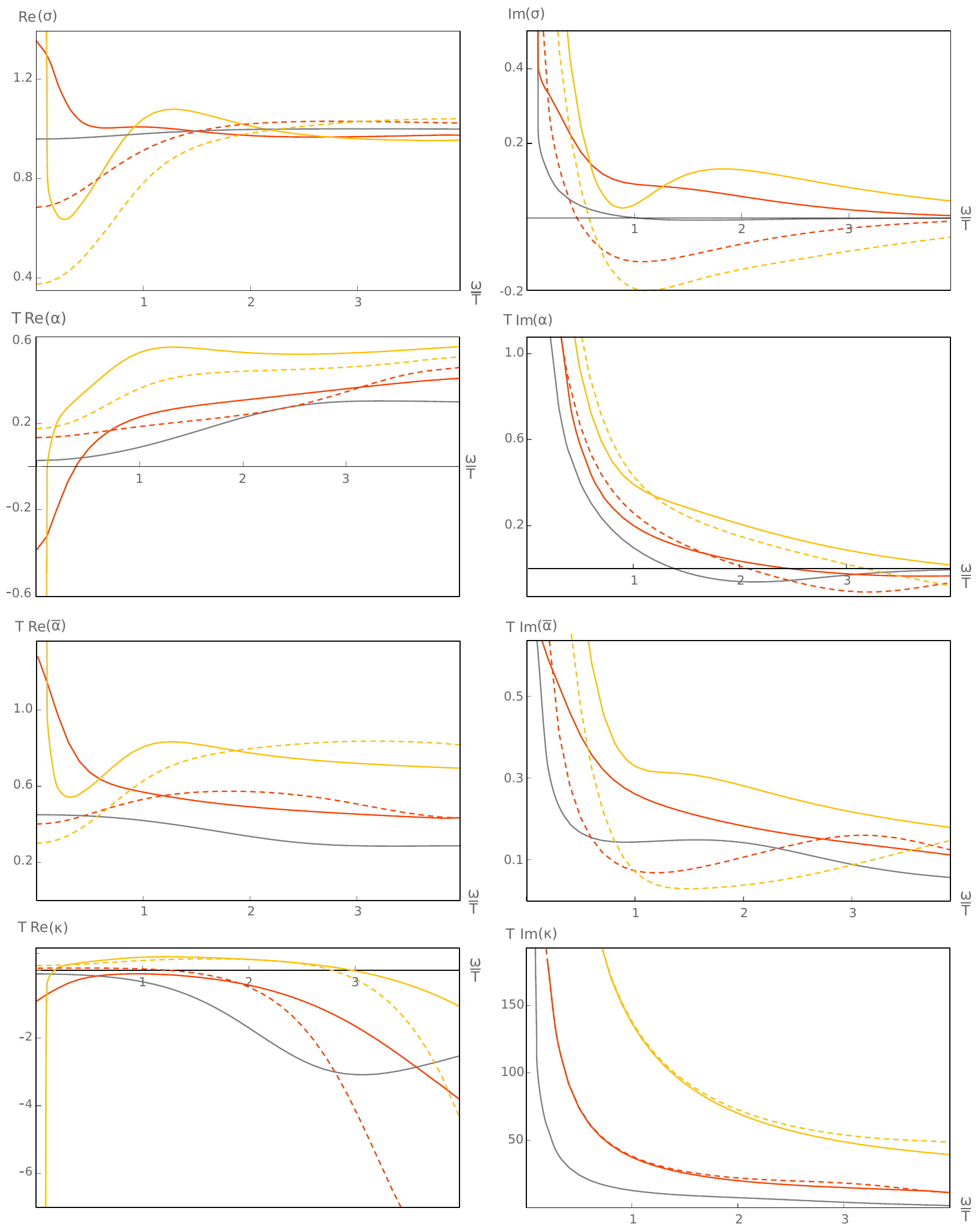}
\caption{\label{fig:ACconductivities1} Conductivities in the  nematic phase for  $T/m_*=0.0408$ (yellow) and  $T/m_*=0.0630$ (orange) and in the Lifshitz phase (solid grey) with $T/m_*=0.0728$, at fixed $\mu/m_*=0.2$. The solid (dashed) lines correspond to $x_+=x+ y$ ($x_-=x-y$) directions. Left and right panels correspond to the real and imaginary parts of the conductivities as a function of $\omega/T$. 
 }
\end{center}
\end{figure*}

The imaginary part of electrical conductivity $\sigma$ at finite chemical potential shows an important qualitative difference from the results  at $\mu=0$, as the latter exhibits a pole at zero frequency, $\Im(\sigma)\approx 1/\omega$. This pole translates into a $\delta-$function in the real part of the electrical conductivity via the Kramers-Kronig relations, which is a feature that our numerical analysis cannot capture.  On the other hand, the continuous part in the limit  $\omega\to 0$ defines $\sigma_{DC}$ in each direction as
\begin{equation}
    \sigma (\omega)\approx \sigma_\delta\, \delta(\omega) + \frac{i \sigma_i}{\omega}+\sigma_{DC}+ {\cal O}(\omega). 
\end{equation}
{Here $\sigma_\delta$ is a divergent contribution that comes from the fact that no momentum dissipation is featured in the model. On the other hand, $\sigma_{DC}$ is the incoherent part of the electrical conductivity that is not associated with momentum dissipation. In a weakly coupled picture it would be related to the momentum conserving scattering of fermions and holes. Even though such picture is lost at strong coupling, the conserved current propagators are still  natural objects to study. Furthermore, as we here  do not consider momentum dissipation, to characterize our model,  our focus is  exclusively on $\sigma_{DC}$.} In  Figs.~\ref{fig:DCconductivity} and ~\ref{fig:ACconductivities1} we show explicitly the behavior of the electrical DC part ($\omega\to 0$) and the finite-frequency conductivities, respectively.  

The DC conductivity in the nematic phase exhibits an anisotropic behavior when  the chemical potential is tuned at a low enough  temperature, as explicitly shown in Fig.~\ref{fig:DCconductivity}(a). Notably, the conductivities along the two principal directions show a non-monotonic behavior as functions of the chemical potential. This is in contrast to 
the  temperature-dependent behavior  at fixed chemical potential, see Fig.~\ref{fig:DCconductivity}(b).  In the low temperature limit  the conductivity shows a very strong anisotropy, as expected in the nematic phase.
As the temperature or the chemical potential approach the critical value for the transition to the Lifshitz phase, the conductivity becomes isotropic. 

Considering   now the full dependence of the conductivities with the frequency, we notice that in  one direction  a Drude-like peak in the electrical conductivity $\sigma$ appears, while  the orthogonal one displays  a depletion,  see Fig. \ref{fig:ACconductivities1}. As the temperature  increases, the electrical conductivity smoothens up, and  the Drude weight thereby diminishes. 
In the direction developing the Drude-like peak, the conductivity also features stronger oscillations as a function of the frequency. A possible explanation for these oscillations could be the appearance of poles in the current-current correlator close to the real $\omega$ axis (long lived excitations). As the temperature increases the poles approach the imaginary $\omega$ axis getting further away from the real $\omega$ axis, therefore displaying  the so called zipper mechanism \cite{Sachdev:2012qgh,Ren:2021rhx}. Hence in our model the finite-frequency electrical conductivity seems to be controlled by the pole structure of the corresponding correlators hiding any imprint of the intermediate Lifshitz behavior shown by the geometry. Regarding the thermal conductivity $\kappa$, its imaginary part remains almost isotropic in the nematic phase, while its real part deviates from isotropy at high and low frequencies.



\section{Conclusions and outlook}
In this paper, we show that the finite density can  tune the quantum phase transition between the nematic and the symmetric phases, as shown in Fig.~\ref{fig:phase-diagram}.  Furthermore,  the optical conductivity    can distinguish the two phases through the explicit anisotropy in the conductivity featured  in the nematic state, see Figs.~\ref{fig:DCconductivity} and \ref{fig:ACconductivities1}. In particular, as the system enters the symmetric phase from the nematic side, the conductivity becomes fully isotropic.

We notice that for a large chemical potential the geometry in our holographic model should approximate that of a charged black hole, while in the low-temperature limit  the deep IR corresponds to an AdS$_2\times R_2$ geometry, as we find by studying the behavior of the entropy with respect to the temperature. The latter geometry is well known to render unstable a charged operator \cite{Gubser:2008px} with the end-point of such instability being a superconducting state \cite{Hartnoll:2008vx}. This possible emergence of the superconducting states from the insulating ones might be a  robust feature of our holographic model, as it may depend only on its global geometric features. This  problem is left for  future investigation. 

As a final remark,  it would be interesting to study constructions from string theory, as they should give further control of the dual field theory considered here. Furthermore, the matter content of our model, which is rather minimal, allows for various directions to extend this pursuit. One particular way would be to mimic the constructions for stringy p-wave superfluids \cite{Ammon:2009fe}. In this respect, it is particularly relevant  that  in the context of holographic Weyl semimetals an extra control over the fermionic sector is obtained  when considering  constructions from string theory~\cite{Fadafan:2020fod}.  

\section*{Acknowledgments:} I.S.L.  would like to thank Samuele Giuli for insightful discussions.  This work was supported by ANID/ACT210100 (R.S.-G. and V.J), the Swedish Research Council Grant No. VR 2019-04735 (V.J.), Fondecyt (Chile) Grant No. 1200399 (R.S.-G.) and No. 1230933 (V.J.), the CONICET grants PIP-2017-1109 and PUE 084 ``B\'usqueda de Nueva F\'isica”, and by UNLP grants PID-X931 (I.S.L. and N.E.G.).  I.S.L. thanks ICTP and Universidad Católica de Chile for hospitality during different stages of this project. I.S.L would like to acknowledge support from the ICTP through the Associates Programme (2023-
2028). N.E.G. thanks Universidad Católica de Chile, Universidad de Concepción and Centro de Estudios Científicos for hospitality during this project.

\bibliographystyle{JHEP}
\bibliography{references}

\end{document}